# Scraping and Clustering Techniques for the Characterization of LinkedIn Profiles

Kais Dai[1], Celia Gónzalez Nespereira[1], Ana Fernández Vilas[1] and Rebeca P. Díaz Redondo[1]
[1]Information & Computing Laboratory,
AtlantTIC Research Center for Information and Communication Technologies-
University of Vigo, 36310, Spain
{kais,celia,avilas,rebeca}@det.uvigo.es

### *Abstract*

*The socialization of the web has undertaken a new dimension after the emergence of the Online Social Networks (OSN) concept. The fact that each Internet user becomes a potential content creator entails managing a big amount of data. This paper explores the most popular professional OSN: LinkedIn. A scraping technique was implemented to get around 5 Million public profiles. The application of natural language processing techniques (NLP) to classify the educational background and to cluster the professional background of the collected profiles led us to provide some insights about this OSN's users and to evaluate the relationships between educational degrees and professional careers.*

## 1. Introduction

The influence of Online Social Networks (OSNs) [1,2,3] on our daily lives is nowadays deeper according to the amount and quality of data, the number of users, and also to the technology enhancement, especially related to the concurrence in the smartphones' market. In this sense, companies but also researchers are attracted by the cyberspace's data and the huge variety of different challenges that the analysis of the data collected from social media opens: capturing public opinion, identifying communities and organizations, detecting trends or obtaining predictions in whatever area a big amount of user-provided data is available.

Taking professional careers as focus area, LinkedIn is undoubtedly one of these massive repositories of data. With 300 million subscribers announced in April 2014 [4], LinkedIn is the most popular OSN for professionals [5], and it is distinctly known as a powerful professional networking tool that enables its users to display their curricular information and to establish connections with other professionals. Given this huge amount of valuable data about education and professional careers, it is possible to explore it for capturing successful profiles, identifying professional groups, and more ambitiously detecting trends in education and professional careers and even predicting how successful a LinkedIn user is going to be in their near future career.

Several web data collection techniques were developed especially related to OSNs. First, OSN APIs ("Application Programming Interface") [6], a win-win relationship between Web Services providers and applications, have emerged in the most popular OSNs such as Facebook, Twitter, LinkedIn, etc. OSN APIs enable the connection with specific endpoints and obtaining encapsulated data mostly by proceeding on users' behalf. Some considerable constraints when following this method arise which impact directly in the amount of collected data: the need of users' authorizations, the demanding processing and storage resources but especially, and in almost of the cases, the data access limitations. For instance, the Facebook Platform Policy only allows to obtain partial information from profiles and do not allow to exceed 100M API calls per day1. In LinkedIn, it is only possible to access the information of public profiles or, having a LinkedIn account, it is possible to access to users' private profiles that are related to us at least in the third grade2.

In this context, crawling techniques appear as an alternative. It basically consists in traversing recursively through hyperlinks of a set of webpages and downloading all of them [7]. With the growth of the cyberspace's size, generic crawlers were less adapted to recent challenges and more accurate crawling techniques were proposed such as topic-based crawlers [8], which consists of looking for documents according to their topics. In this sense, a more economical alternative, referred to as scrapping, gets only a set of predetermined fields of each visited webpage, which allows tackling not only with the increasingly size of cyberspace but also with the size of the retrieved data. Please, note that crawlers deal generally with small websites, whereas scrappers are originally conceived to deal with the scale of OSNs.

However, obtaining the data is only one part of the problem. Generally, once gathered, the data go through a series of analysis techniques. In this context, clustering techniques were widely applied in order to explore the available data by grouping OSNs' profiles to discover hidden aspects of the data and, on this basis, provide accurate recommendations for users' decision-support. For example, in [9] the authors use crawling methods to study the customers' behaviour in Taobao (the most important e-commerce platform in China). At best, the data to be grouped have some structure which allows to easily define some distance measure but, unfortunately, this is not always the case when users freely define their profiles. Such is the case for LinkedIn users. If so, discover/uncover hidden relationships involves applying some NLP (Natural Language Processing) Techniques.

The main contributions of this paper can be summarized in: (1) obtaining a LinkedIn dataset, which does not exist to our knowledge; and performing exploratory analysis to uncover (2) educational categories and their relative appearance in LinkedIn and (3) professional clusters in LinkedIn by grouping profiles according to the free summaries provided by LinkedIn users.

This paper is organized as follows: Next section provides the background of OSNs crawling strategies, natural language processing and clustering techniques. After briefly introducing LinkedIn (Section 3), we provide a descriptive analysis of the data collection process and the obtained dataset in Section 4. The classification of educational background and clustering of professional background are shown in Sections 5 and 6, respectively. Finally, we draw some conclusions and perspectives in the last section.

## 2. BACKGROUND

Since their emergence, Online Social Networks (OSNs) have been widely investigated due to the exponential growing number of their active users and consequently to their socio-economic impact. In this sense, the challenging issue of collecting data from OSNs was addressed by a multitude of crawling strategies. These strategies were implemented according to the specificities of each social network (topology, privacy restrictions, type of communities, etc.) but also regarding the aim of each study. One of the most known strategies is the Random Walk

[1] https://www.facebook.com/about/privacy/your-info [Last access: 24 October 2014]
[2] https://www.linkedin.com/legal/privacy-policy [Last access: 24 October 2014]

(RW) [10] which is based on a uniform random selection from a given node among its neighbours [11]. Other techniques, such as the Metropolis-Hasting Random Walk (MHRW) [12], Breadth First Search (BFS) [11], Depth First Search (DFS) [13], Snowball Sampling [11] and the Forest Fire [14] were also presented as graph traversal techniques and applied to address OSNs crawling issues. One of the aims of crawling strategies is profiles retrieval. After that, there are some techniques that allow us to match the different profiles of one user in the different OSNs [14].

As pointed out in the introduction, dealing with textual data leads inevitably to the application of NLP (Natural Language Processing) techniques which mainly depends on the language level we are dealing with.(pragmatic, semantic, syntax) [15]. A good review of all NLP techniques could be found in [16].

In our case, we have centred in the application of NLP techniques to Information Retrieval (IR) [17]. The main techniques in this field are: (i) Tokenization, that consists in divide the text into tokens (keywords) by using white space or punctuation characters as delimiters [18]; (ii) Stop-words removal [19], which lies in removing the words that are not influential in the semantic analysis, like articles, prepositions, conjunctions, etc.; (iii) Stemming, whose objective is mapping words to their base form ([20] presents some of the most important Stemming algorithms); (iv) Case-folding [18], which consists in reduce all letters to lowercase, in order to be able to compare two words that are the same, but one of them has some uppercase. Applying these NLP techniques allow us to obtain a Document-Term Matrix, a matrix that represent the number of occurrences of each term in each document.

The main objective of obtaining the Document-Term Matrix is to apply classification and clustering methods [21] over the matrix, in order to classify the user's profiles. Classification is a supervised technique, whose objective is grouping the objects into predefined classes. Clustering, on the contrary, is an unsupervised technique without a predefined classification. The objective of clustering is to group the data into clusters in base of how near (how similar) they are.

There are different techniques of clustering [22], which can mainly be labelled as partitional or hierarchical. The former obtains a single partition of the dataset. The latter obtains a dendrogram (rooted tree): a clustering structure that represents the nested grouping of patterns and similarity levels at which groupings change.

K-means [23] is the most popular and efficient partitional clustering algorithm and has important computational advantages for large dataset, as Lingras and Huang show in [24]. One of the problems of the K-means algorithm is that it is needed to establish the number of clusters in advance. Finally, the gap statistic method [25] allows estimating the optimal number of clusters in a set of data. This method is based on looking for the Elbow point [26], or the point where the graph that represents the number of cluster versus the percentage of variance explained by clusters starts to rise slower, giving an angle in the graph.

## 3. SUMMARY OF THE LINKEDIN EXPERIMENT

The primary contribution of this paper is obtaining an anonymized dataset by scrapping the public profile of LinkedIn users subject to the terms of LinkedIn Privacy Policy. LinkedIn public profile refers to the user's information that appears in a Web Page when the browser is external to LinkedIn (not logged in). Although users can hide their public profile to external search, it is exceptional in this professional OSN. In fact, what users normally do is hiding some sections. The profile's part in Fig. 1.(a) includes personal information and current position. To respect LinkedIn privacy policy, we discarded that part for the study and we anonymized profiles by unique identifiers. The profile's part in Fig. 1.(b) contains information about educational and professional aspects of the user, being the most relevant data for our exploratory analysis of LinkedIn.

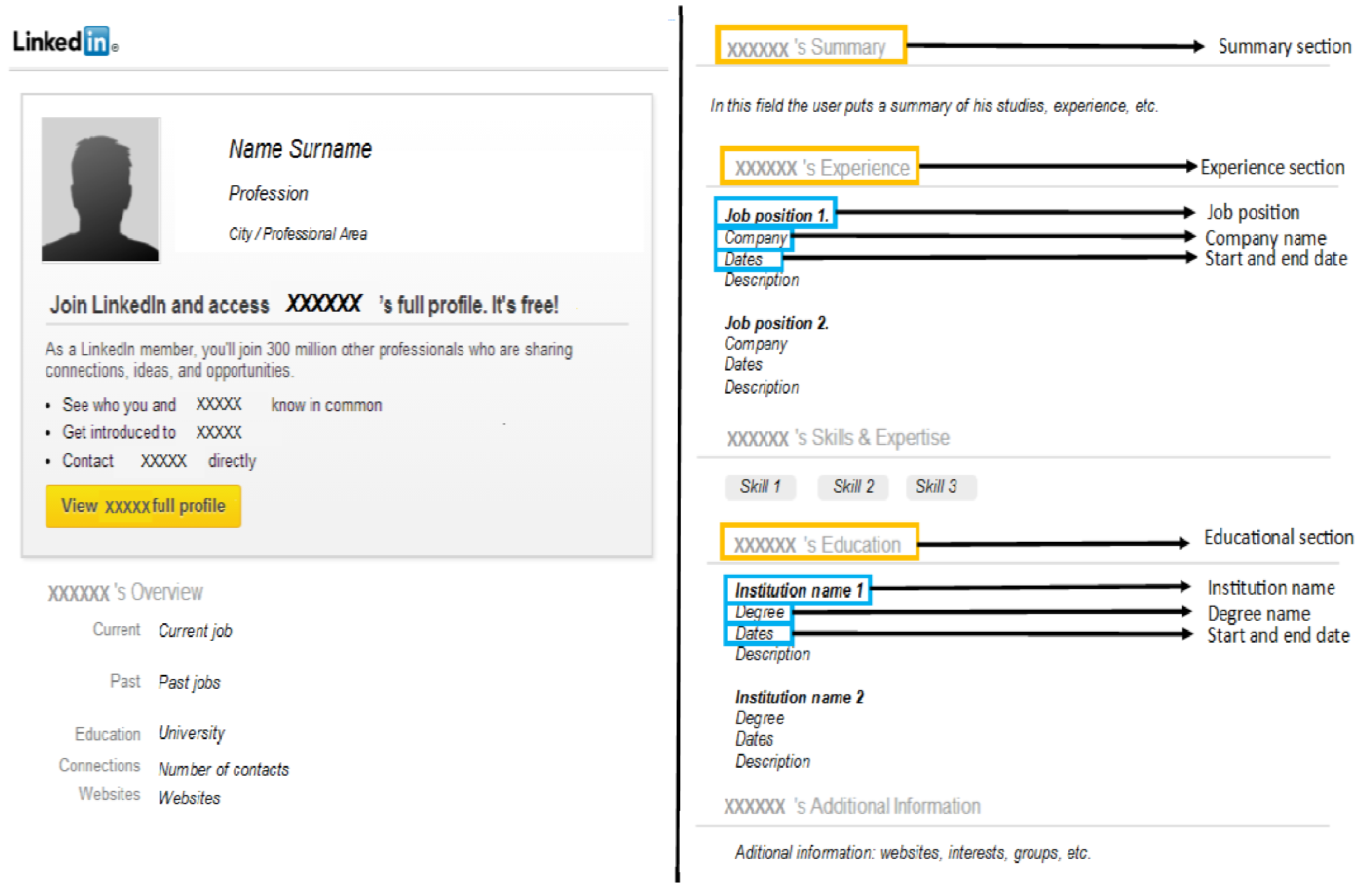

(a) (b)

Fig. 1. Example of a LinkedIn public profile

In this analysis, we grouped profiles according to the former dimensions in order to uncover potential relationships between groups. This grouping process may be done applying different techniques. As there is an established consensus around academic degrees, we apply classification as a supervised learning technique that assigns to each profile an educational level defined in advance. On the other hand, as this consensus is far to be present in the job market, we apply clustering as an unsupervised technique which groups profiles according to their professional similarity.

First, the experiment inspects the educational background to give response to a simple question: What academic backgrounds are exhibited for LinkedIn users? From that, we classify the educational levels in advance and assign each profile to its higher level by processing the text in the educational section of the profile (see Fig. 1.(b)).

Second, the experiment also inspects the professional background. For this, there is not a wide consensus in professionals' taxonomy or catalogue for professional areas or professional levels. Not having so in advance, clustering turns into an appropriate unsupervised technique to automatically establish the groups of professional profiles which are highly similar to each other (according to the content in the Summary (natural text) and Experience (structured text) sections of the profile (see Fig. 1.(b)).

## 4. OBTAINING A DATASET FROM LINKEDIN

One way to get users' profiles from LinkedIn is to take advantage of its APIs (mainly by using the JavaScript API which operates as a bridge between the users' browser and the REST API and it allows to get JSON responses to source's invocations) by using the "field selectors" syntax. In this sense, and according to the user's permission grant, it is possible to get information such as the actual job position, industry, three current and three past positions, summary description, degrees (degree name, starting date, ending date, school name, field of study, activities, etc.), number of connections, publications, patents, languages, skills, certifications, recommendations, etc. However, this method depends on the permission access type (basic profile, full profile, contact information, etc.) and on the number of granted accesses. Thus, this data collection method is not the most appropriate for our approach, since a high number of profiles are needed in order to overcome profiles incompleteness.

Consequently, a scraping strategy was elaborated in order to get the maximum number of public profiles in a minimum of time. LinkedIn provides a public members' directory, a hierarchy taking into account the alphanumeric order[3] (starting from the seed and reaching leafs that represent the public profiles). In this sense, we applied a crawling technique to this directory using the Scrapy framework on Python[4] based on a random walk strategy within the alphabetic hierarchy of the LinkedIn member directory. Our technique relies on HTTP requests following an alphabetic order to explore the different levels starting from the seed and by reaching leafs which represent public profiles. During the exploitation phase, we dealt with regular expressions corresponding to generic XPaths[5] that look into HTML code standing for each public profile and extract required items.

## 4.1. *Data collection process*

During the exploration phase, hyperlinks of the directory's webpages (may be another intermediate level of the directory or simply a public profile) are recursively traversed and added to a queue. Then, the selection of the next links to be visited is performed according to a randomness parameter. At each step, this solution checks if the actual webpage is a leaf (public profile) or only one of the hierarchy's levels. In the former case, the exploitation phase starts and consists of looking for predetermined fields according to their HTML code and extracting them to build up the dataset. Exploitation puts special emphasis in educational and professional backgrounds of public users' profiles, so it extracts the following fields: last three educational degrees, all actual and previous job positions, and the summary description. Table 1 shows a summary of the extracted data and their relations with the sections of the user's profile web page (Fig. 1). As it is shown in the following sections, the deployed technique showed its effectiveness in terms of the number of collected profiles.

Table 1. Description of the extracted fields

| **Field name** | **Field in user profile** | **Description** |
|---|---|---|
| *postions_Overview* | Job positions (Experience section) | All actual and previous job positions. |
| *summary_Description* | Summary section. | Summary description. |
| *education_Degree1* | Degree (Educational section) | Last degree. |
| *education_Degree2* | Degree (Educational section) | Next-to-last degree. |
| *education_Degree3* | Degree (Educational section) | Third from last degree. |

## 4.2. *Filtering the Dataset*

After obtaining the dataset, we apply some filters to deal with special features of LinkedIn. First, according to their privacy and confidentiality settings, LinkedIn public profiles are slightly different from one user to another. Some users choose to display their information publicly (as it is the default privacy setting), others partially or totally not. This feature mainly impact on the completion level of profiles in the dataset, since the collected information obviously depends on its availability for an external viewer (logged off). To tackle this issue, we filter the original dataset (after scrapping) by only considering profiles with some professional experience and at least one educational degree. Second, LinkedIn currently support more than twenty languages from English to traditional Chinese. Users take advantage of this multilingual platform and fulfil their profiles information with their preferred language(s). Although multilingual support is part of our future work, we filter the dataset by only considering profiles written in English.

[3] http://www.linkedin.com/directory/people-a [Last access: 24 October 2014]
[4] http://scrapy.org [Last access: 24 October 2014]
[5] Used to navigate through elements and attributes in an XML document.

### 4.3. *Description of the data set*

As aforementioned, personal information such as users' names, surnames, locations, etc. are not scraped from public profiles. Thus, the collected data is anonymized and treated with respect to users' integrity. Originally, the total number of scrapped profiles was 5,702.544. After performing the first filter by keeping only profiles that mention at least a "job position" and an "educational degree", the total size of the data set became 920.837 profiles. Another filter is applied in order to get only profiles with a description of positions strictly in English. Finally, 217.390 profiles composed the dataset. Not all of these profiles have information in all fields. Table 2 shows the number of profiles that have each field.

Table 2. Profiles number with complete fields

| **Field name** | **Number of profiles** |
|---|---|
| *postions_Overview* | 217.390 |
| *Summary_Description* | 84.781 |
| *education_Degree1* | 205.595 |
| *education_Degree2* | 127.764 |
| *education_Degree3* | 45.002 |

## 5. CLASSIFICATION OF THE EDUCATIONAL BACKGROUND

As described in the previous section, after filtering, anonymizing and pre-processing the data, our dataset retains positions and degrees (with starting and ending dates) and the free summary if available. Despite of this fact, we merely rely on degree fields to establish the educational background. Probably, the profile's summary may literally refer to theses degrees but taking into account the results in section 4.3, we can consider that degree fields are included even in the most basic profiles.

Unfortunately, a simple inspection of our dataset uncovers some problems related with degrees harmonization across different countries. That is, the same degree's title may be used in different countries for representing different educational levels. For this particular reason we opted for a semi-automated categorization of the degree's titles. The UNESCO6's ICSED7 education degree levels classification presents a revision of the ISCED 1997 levels of education classification. It also introduces a related classification of educational attainment levels based on recognised educational qualifications. Regarding this standard and the data we have collected from LinkedIn, we can focus on four different levels which are defined in Table 3.

- **PhD or equivalent (level 8):** This level gathers the profiles that contain terms such as Ph.D. (with punctuation), PHD (without punctuation), doctorate, etc. As it is considered as the highest level, we have not any conflict with the other levels (profiles which belongs to this category may have other degrees title as Bachelor, Master, etc.) and we do not need to apply any constraint.

[6] United Nations Educational Scientific and Cultural Organization.
[7] Institute for Statistics of the United Nations Educational, Scientific and Cultural Organization (UNESCO): "International Standard Classification of Education: ISCED 2011" (http://www.uis.unesco.org/Education/Documents/isced-2011-en.pdf), 2011.

- **Master or equivalent (level 7):** This level includes the profiles which contain, in their degrees' fields description, one of the related terms evoking master's degree, engineering, etc. Profiles belonging to this category do not have to figure in the previous section so we obtain profiles with a master degree or equivalent as the highest obtained degree (according to LinkedIn users' description).
- **Bachelor or equivalent (level 6):** contains a selection of terms such as bachelor, license and so on. So obviously profiles here do not belong to any of the highest categories.
- **Secondary or equivalent (both levels 5 and 4):** contains LinkedIn profiles with the degrees' titles of secondary school.

Table 3. Keywords of the 4-levels classification.

| **Level** | **Keywords** |
|---|---|
| PhD | *phd ph.d. dr. doctor dottorato doctoral drs dr.s juris pharmd pharm.d dds d.ds dmd postdoc resident doctoraal edd* |
| Master | *master masters msc mba postgraduate llm meng mphil mpa dea mca mdiv mtech mag Maîtrise maitrise master's mcom msw pmp dess pgse cpa mfa emba pgd pgdm masterclass mat msed msg postgrad postgrado mpm mts* |
| Bachelor | *bachelor bachelors bsc b.sc. b.sc licentiate bba b.b.a bcom b.com hnd laurea license licenciatura undergraduate technician bts des bsn deug license btech b.tech llb aas dut hbo bpharm b.pharm bsba bacharel bschons mbbs licenciada bca b.ca bce b.ce licenciado bachiller bcomm b.comm bsee bsee cpge bsw b.sw cess bachillerato bas bcs bcomhons bachalor bachlor bechelor becon bcompt bds bec mbchb licencjat bee bsme bsms bbs graduado prepa graduat technicians technicien tecnico undergrad bvsc bth bacharelado* |
| Secondary | *secondary sslc ssc hsc baccalaureate bac dec gcse mbo preuniversity hnc kcse ssce studentereksamen secondaire secundair igcse ossd vmbo htx* |

Using the keywords in Table 3 as a criteria for the classification over the filtered data set, we obtain the distribution of levels in Fig. 2.

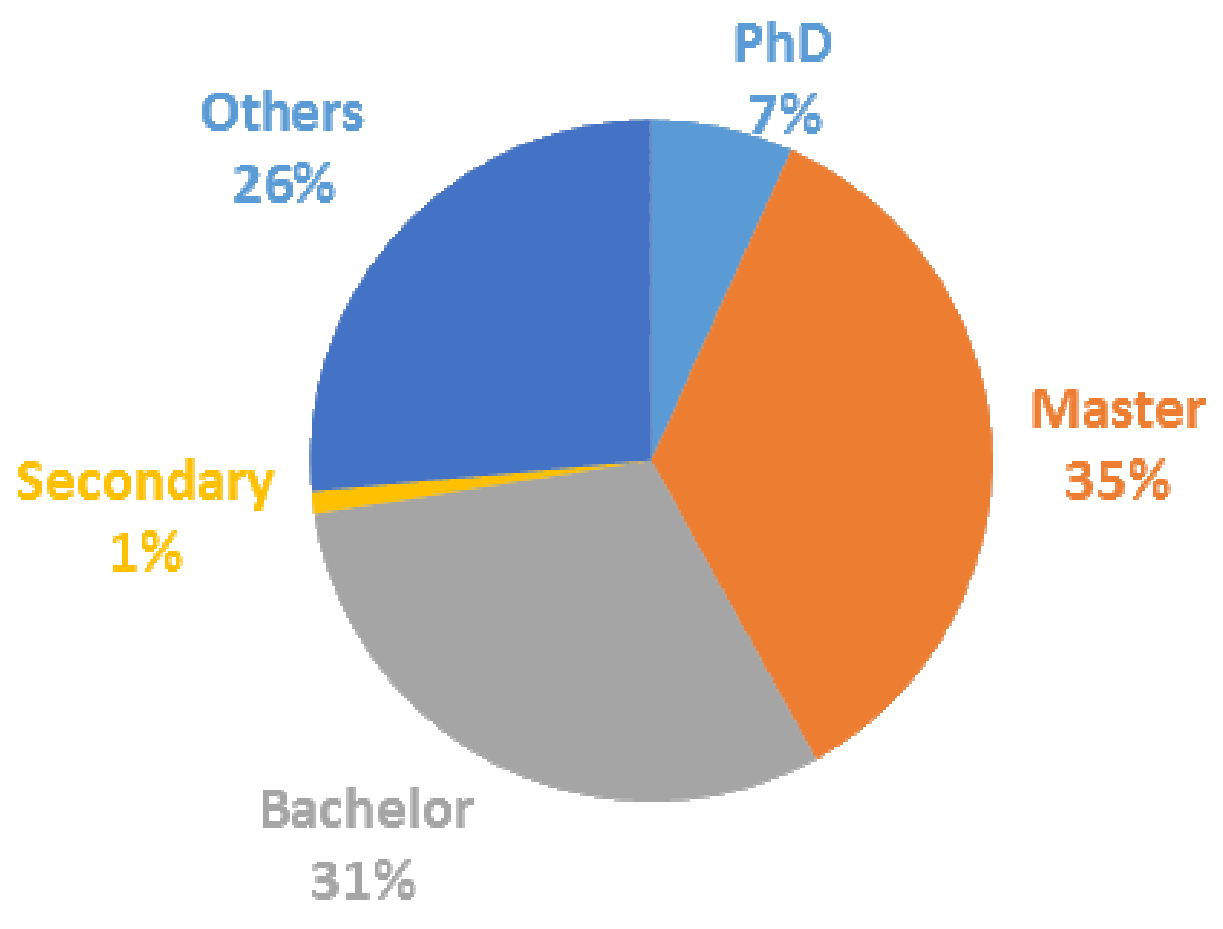

Fig. 2. Classification of the educational background of our LinkedIn data set.

## 6. CLUSTERING THE PROFESSIONAL BACKGROUND

The purpose of this section is to analyse the professional background of the collected profiles. As mostly the rest of the fields of a typical LinkedIn profile, current and past job positions but also user's summary description can be managed freely and without special use of predefined items. In this sense, LinkedIn profile's fields are not conform to a specific standard (as the UNESCO one used for the classification of users according to their educational information). With the freedom given to users to present their job positions and especially their summaries' description, the possible adoption of the classification approach is clearly more difficult. Also, apart from the fact that some professional profiles are more multidisciplinary then others, we strongly believe that each professional profile is proper, and this by considering the career path as a whole. In this sense, we consider that applying clustering is more appropriate in this case. But before doing that, we must apply some transformations to our refined dataset.

### 6.1. Text Mining

In this context, we have to deal with a "corpus" which is defined as "a collection of written texts, especially the entire works of a particular author or a body of writing on a particular subject,[…], assembled for the purpose of linguistic research"8. In order to perform text mining operations on the data set, we need to build a corpus using the set of profile fields we are interested in. In this sense, the job positions and summary description of all profiles of the refined data set will be used to construct the so called corpus. Furthermore, a series of NLP functions must be applied on the corpus in order to build the Document Term Matrix (DTM) which represents the correspondence of the number of occurrence of each term (stands for a column) composing the corpus to each profile description (a document for each row).

First, the transformation of the corpus' content to lowercase is performed for terms comparison issues. Then, stop-words (such as: "too", "again", "also", "will", "via", "about", etc.), punctuations, and numbers are removed. Finally, white spaces are stripped in order to avoid some terms' anomalies. Stemming the corpus is avoided in this work because it didn't demonstrate better tokenization results with this data set.

The generation of the DTM can now be performed using the transformed corpus. With a 217.390 document, this DTM has high dimensions especially regarding the number of obtained terms. Analysing such a data structure become memory and run time consuming: we are dealing with the so called curse of dimensionality problem [27]. In order to tackle the latter, we opted to push our study forward and focus on profiles which belongs to only one educational category. As being the highest level, and with 14.650 profiles, the 8th category (PhD) seems to be the ideal candidate for this analysis.

In this sense, we subsetted the profiles of this category from the data set and applied the same NLP transformations described earlier in this section to a new corpus. Then, we generated the DTM by only considering terms' length more than 4 letters (simply because we get better results while inspecting the DTM terms). In fact, with its 14.650 document, this DTM is composed of 75.624 terms and it is fully sparse (100%). In such cases (high number of terms), the DTM may be reduced by discarding the terms which appears less than a predefined number of times in all DTM's documents (correspond to the DTM's columns sum) or by applying the TF-IDF (Term Frequency-Inverse Document Frequency) technique as described in [28]. By making a series of tests over our 14.650 document's DTM, 50 seemed to be the ideal threshold of occurrences' sum among all documents. After filtering the DTM, we obtained a matrix with only 1.338 term, which will be more manageable within the next steps of this analysis.

---

[8] Definition of "Corpus" in Oxford dictionaries, (British & World English), *http://www.oxforddictionaries.com.* [Last access: 24 October 2014].

### 6.2. Clustering

Considering the size of the obtained DTM and as the efficiency and performance of K-means were widely demonstrated by the data mining community, we opted for the application of a heuristic-guided K-means clustering method. As the predetermined number of clusters "k" is a sensitive parameter for this clustering task (since it guides the algorithm to make the separation between the data points) and in order to address this issue, we applied the elbow method, as a heuristic, to guide the choice of the "k" parameter.

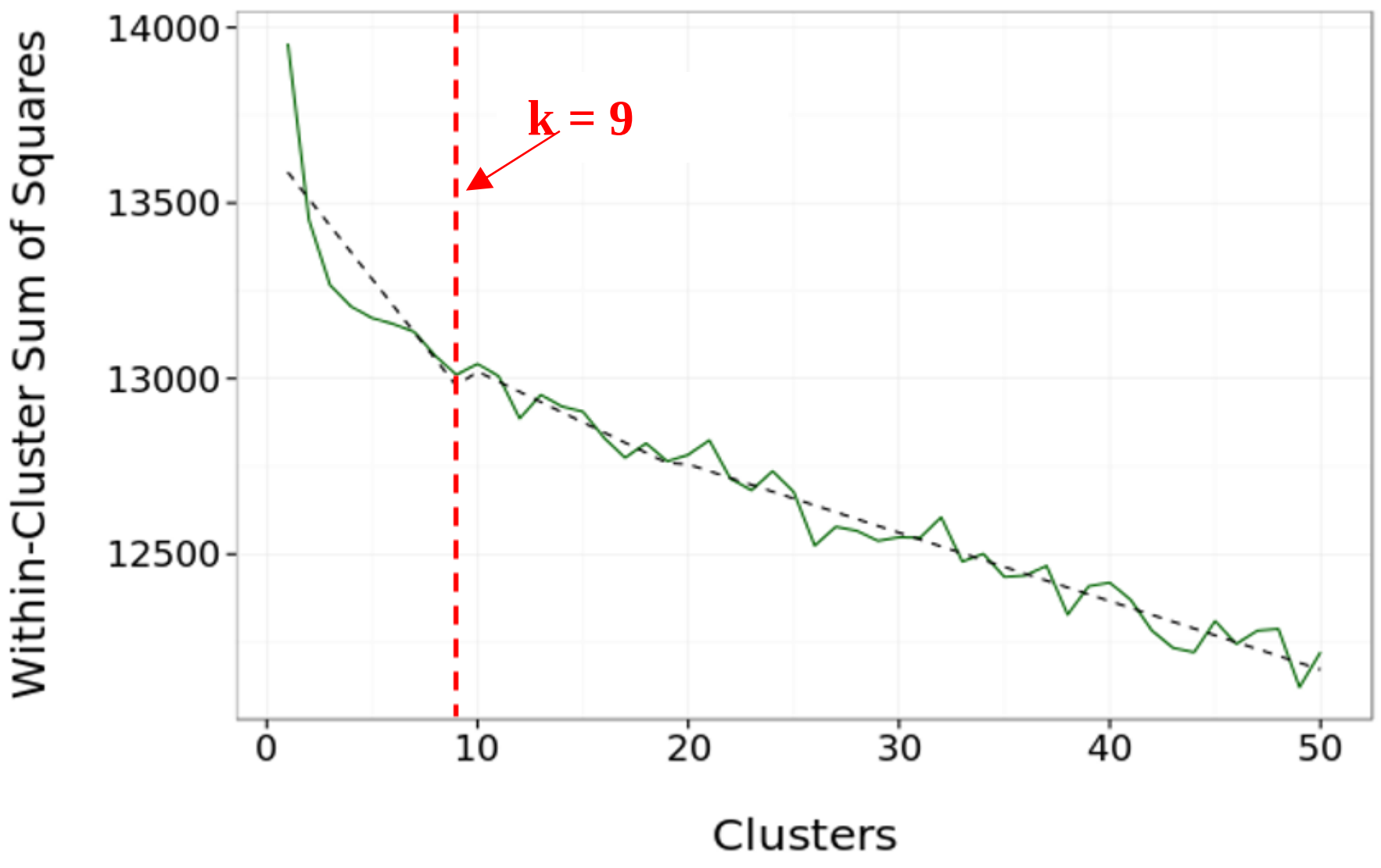

Fig. 3. Reduction in cost for values of "k".

As shown in Figure 3, the elbow method is based on the computation of the squared distance's total for each cluster. With the collection of the k-means' results for different k-values (here from 1 to 50), the analysis of the generated graph allows us to determine the value k that should be considered by taking the breakpoint (marks the elbow point). In our case, and according to Figure 3, we selected 9 clusters.

In contrast with the educational background, classifying profiles according to their professional description become a complex task especially when we have to deal with multidisciplinary profiles. Depending on their positions' history, working field(s), experience, etc., these aspects of the different career paths highlight the singularity of each profile. So, the application of a clustering technique rather than the semi-supervised classification as performed on the educational profiles' description takes all its sense.

Actually, the aim of applying a clustering technique here is to group similar LinkedIn profiles within a restricted category (PhD level) according to their professional background, and thus, to build a methodology for groups' detection. As pointed out earlier, 9 is the appropriate k value to perform k-means clustering. After running the algorithm, we have obtained the clusters distribution in Table 4.

Table I. Number of profiles for each cluster.

| Cluster id | Number of profiles |
|---|---|
| 1 | 199 |
| 2 | 2416 |
| 3 | 210 |
| 4 | 1846 |
| 5 | 374 |
| 6 | 5152 |
| 7 | 981 |
| 8 | 3090 |
| 9 | 75 |

Furthermore, we have decided to analyse the most frequent terms in the professional description of each cluster's profiles. We began by considering only terms associated to profiles belonging to each cluster. Then, we have sorted these terms according to their occurrence's frequency among all profiles. And finally, we have constructed a Tag Cloud for each cluster in order to enable the visualization of these results. Consequently, we have obtained the 9 tag clouds.

The existence of terms co-occurring among different tag clouds can be explained by the fact that there are some common terms used by the majority of profiles of this educational category (PhD level). So, encountering redundant terms such as "university", "professor", "teaching" etc., makes sense when considering this aspect. Indeed, scrutinizing each tag cloud can lead us to the characterization of the profiles' groups. This exercise must be performed regarding all the professional aspects of a profile description (such as the "job position(s)" and "working field(s)", etc.) but also by comparing each tag cloud to all others in order to spotlight its singularity.

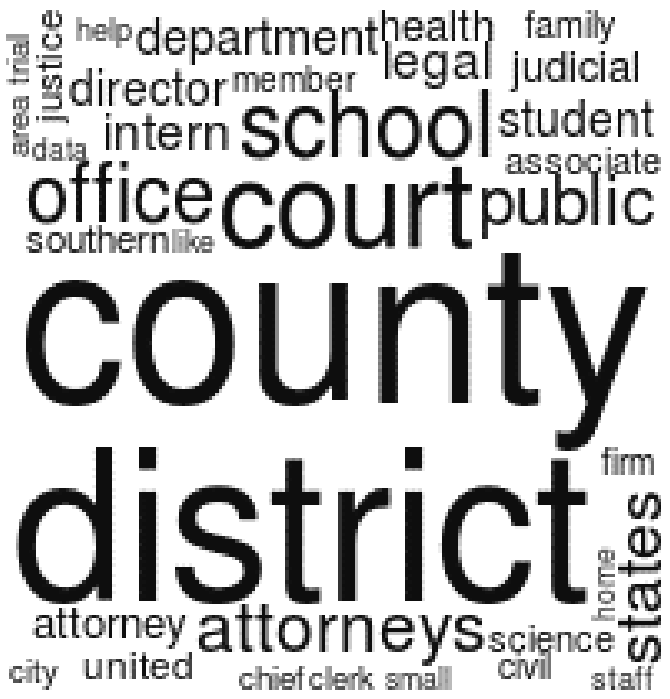

Fig.4. 1. Tag Cloud of cluster "1" profiles.

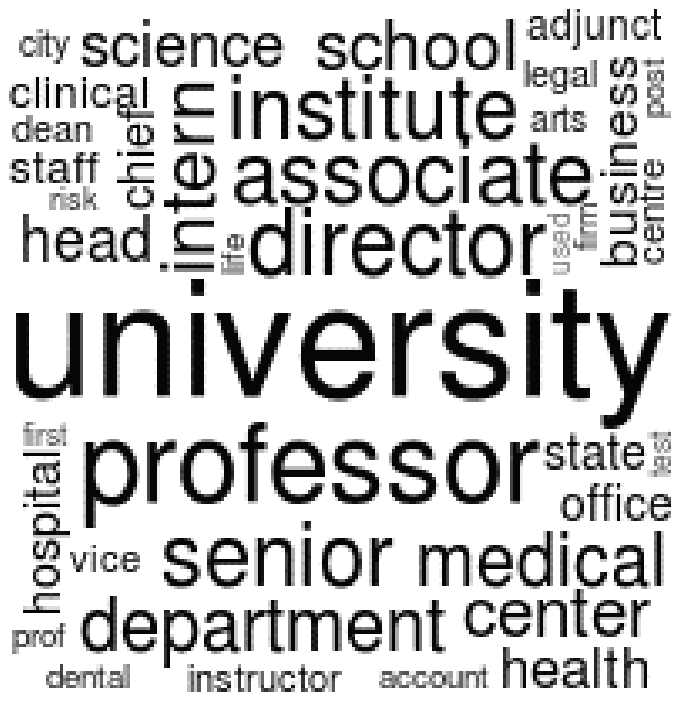

Fig.4. 2. Tag cloud of cluster "2" profiles.

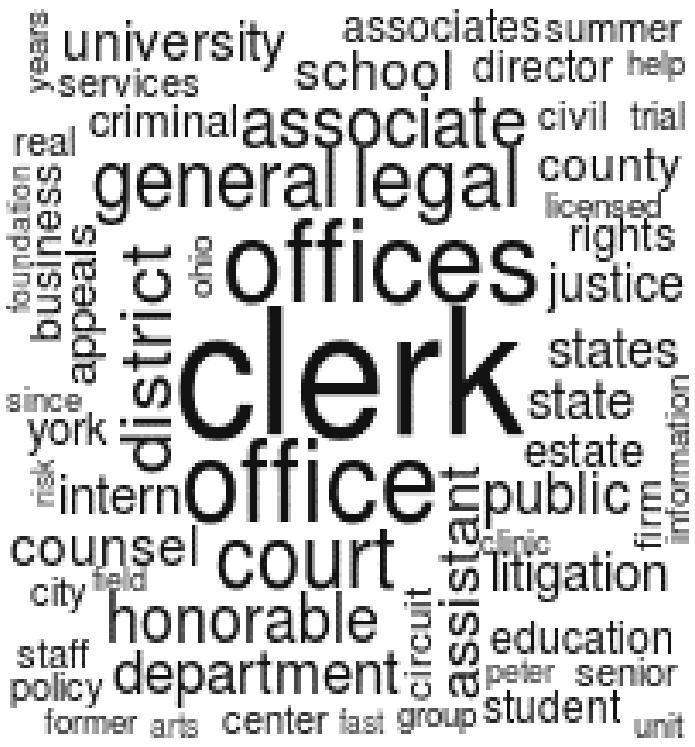

Fig.4. 3. Tag cloud of cluster "3" profiles

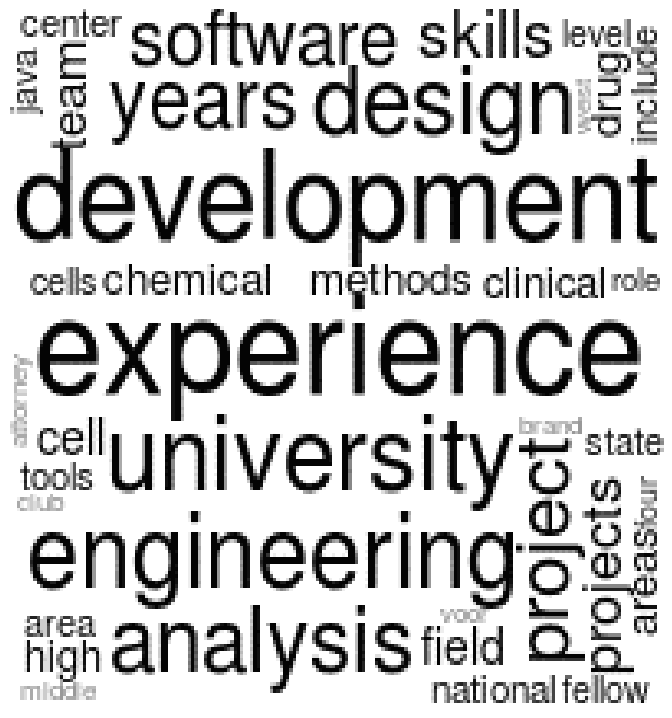

Fig.4. 4. Tag cloud of cluster "4" profiles

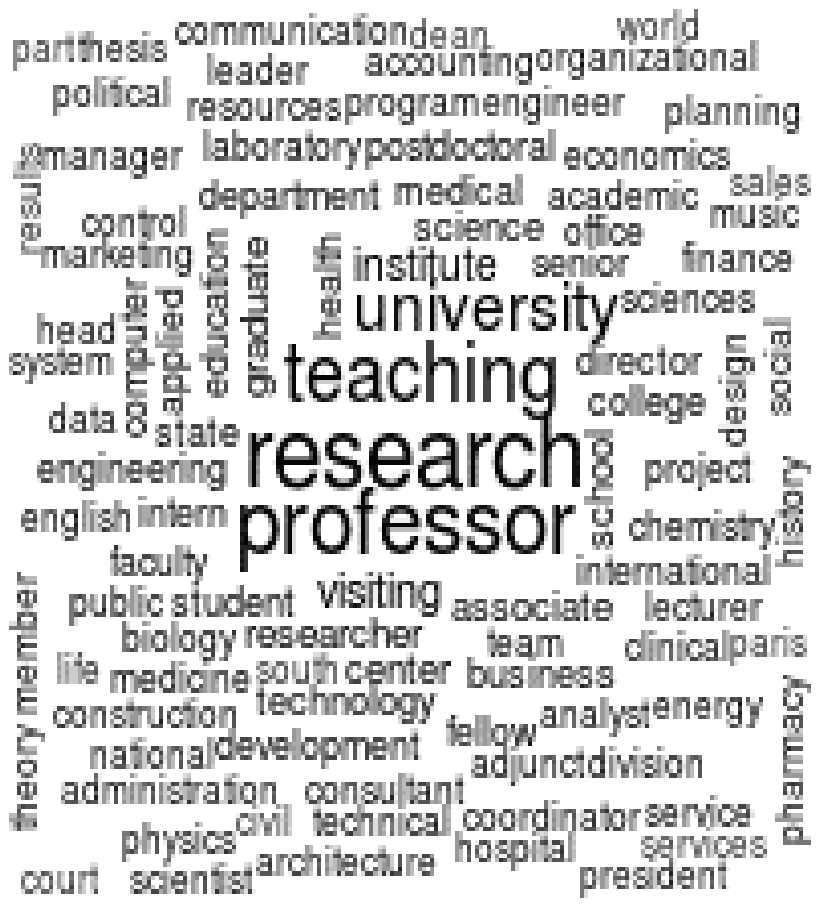

Fig.4. 5. Tag cloud of cluster “5” profiles

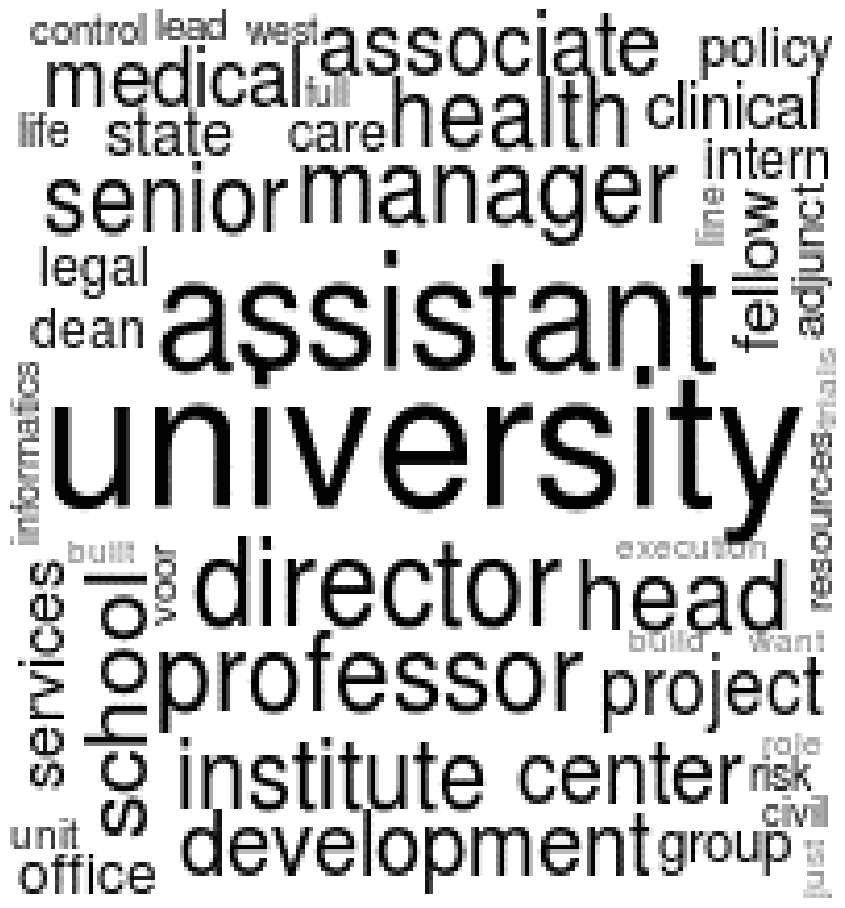

Fig.4. 6. Tag cloud of cluster “6” profiles

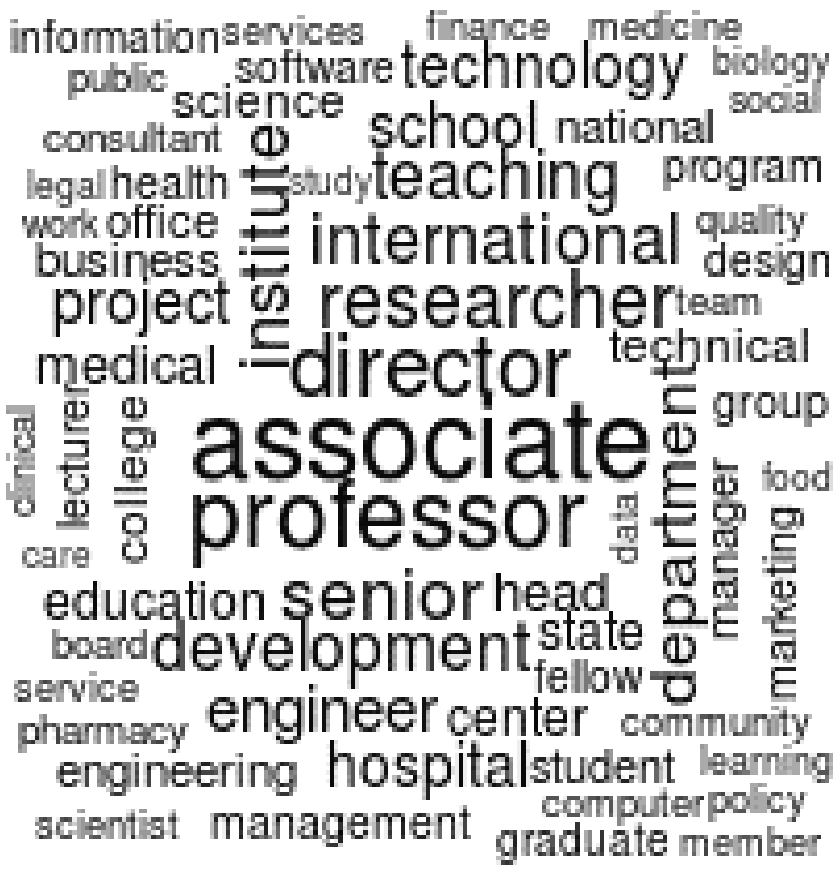

Fig.4. 7. Tag cloud of cluster “7” profiles

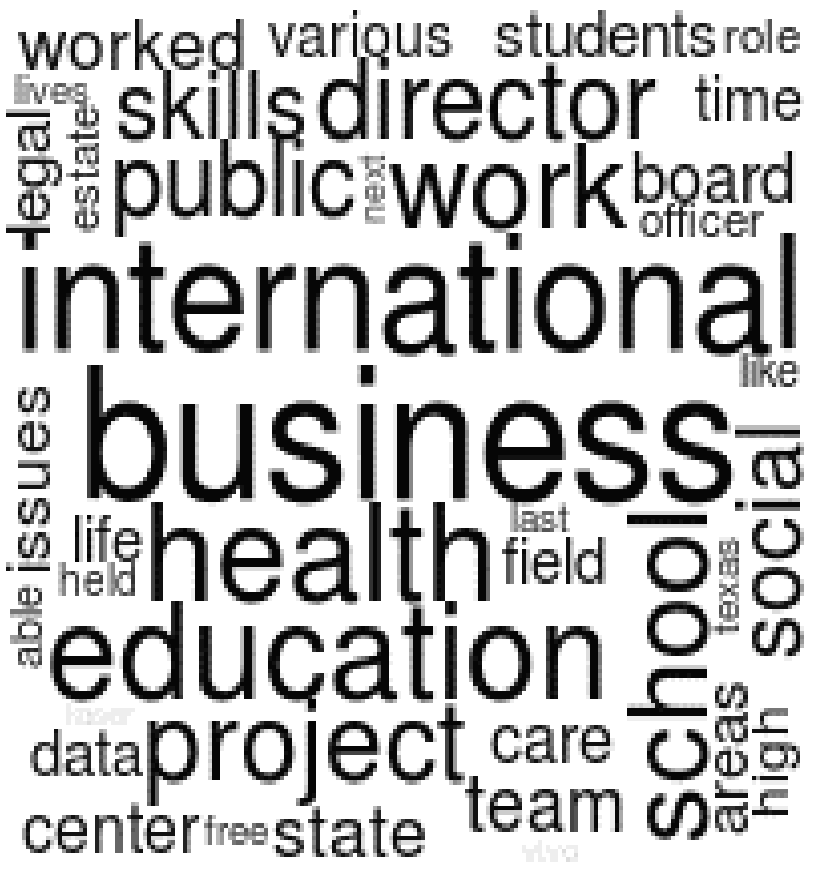

Fig.4. 8. Tag cloud of cluster “8” profiles

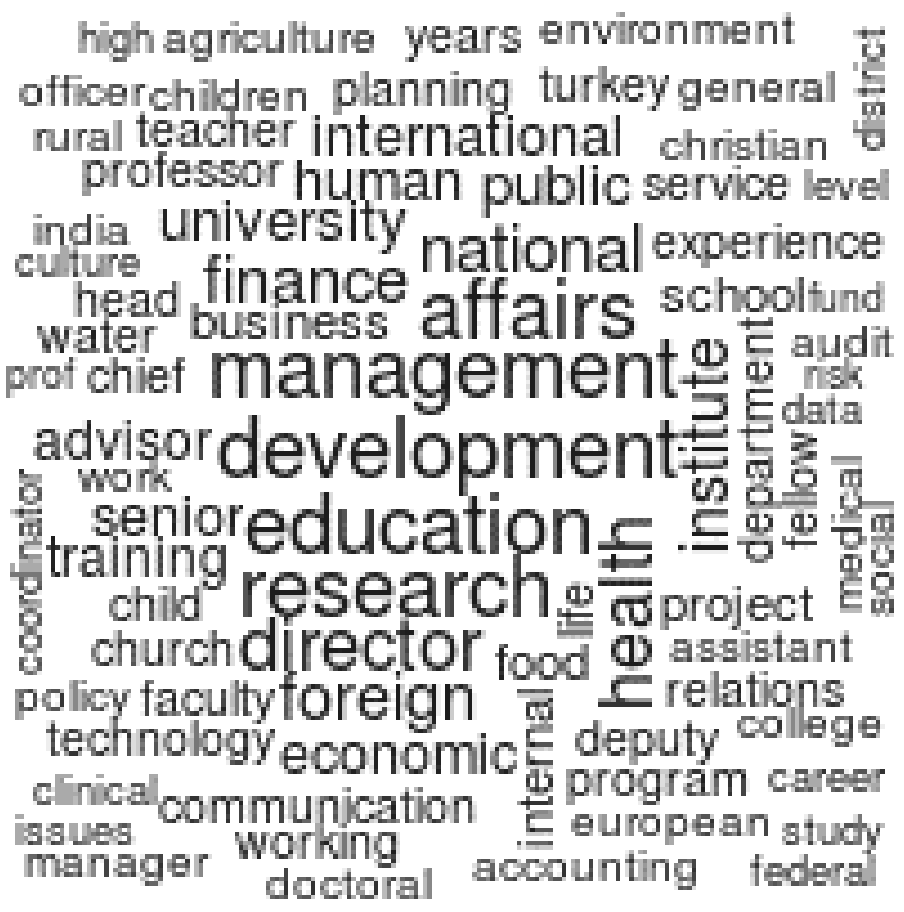

Fig.4. 9. Tag cloud of cluster “9” profiles

Fig. 4. Reduction in cost for values of "k".

## 7. RESULTS

The characterization of the obtained clusters by interpreting their correspondent tag clouds can lead us to draw some conclusions about the different professional groups inside the PhD level category. A scrutiny of the tag clouds lead us to distinguish between these different groups.

Administrative, technical, academic or business are the main discriminative aspects depicted here to characterize these profiles. Also, other major cross-cutting aspects must be considered as the working environment (private or public sector), field (health, technology, etc.) but also the position's level (senior, director, assistant, etc.).

The characterization process takes into account the most relevant terms used in all the profiles (professional description) constituting each cluster. An eventual subtraction of common tags could be conducted but a general interpretation of the tag clouds' results should firstly takes into account the most frequent terms.

Having 199 profiles, cluster number 1 (whose tag cloud is depicted in Fig 4.1) encompasses with the administrative public sector terminology (such as "public", "county", "district, etc.). The profiles of this group are related to the field of legal or juridical science. The second tag cloud describes high level academic researchers by the use of terms such as "university", "professor", "senior", etc.

In contrast with the first one which is also dealing with administrative profiles in the field of legal sciences, the third tag cloud describes professional profiles working in the private sector ("clerk"). The fourth tag cloud describes more technical profiles', characterized via terms such as "development", "engineering", "design", etc.

Represented by its associated tag cloud, cluster 5 clearly spotlight academic profiles which are more involved in teaching experiences ("university", "teaching", "professor", etc.) This aspect is consolidated by the existence of a terminology related to numerous fields of study (economics, technology, medical, etc.).

The sixth cluster represents the most common profiles of the PhD level category (by grouping 5.152 profiles). It is characterized by academic profiles of the range of "assistant professor", "project manager" and they are more research oriented.

Tag cloud number 7 illustrates another group of profiles working in the academic field and mainly composed of "associate professor", which are involved in international projects and may have some administrative responsibilities in their research organizations. The jargon used in the eighth tag cloud clearly describes business profiles working in international settings. Finally, and compared to the previous tag clouds, the last one describes a multidisciplinary profile which takes advantage of all the discussed aspects. With its 75 profiles, the ninth cluster is composed by LinkedIn profiles that encompass different job positions in management, research, teaching, etc.

## 8. DISCUSSION AND FUTURE WORK

Very popular social networks, like Twitter or Facebook, have been intensively studied in the last years and it is reasonable easy to find available datasets. However, since LinkedIn is not as popular, there are not datasets gathering information (profiles and interactions) of this social network. In this paper, we have applied different social mining techniques to obtain our own dataset from LinkedIn which has been used as data source for our study. We consider that our dataset (composed of 5.7 million profiles) is representative of the LinkedIn activity for our study. Few proposals face analysis using this professional social network. In [29] the authors provide a statistical analysis of the professional background of LinkedIn profiles according to the job positions. Our approach also tackles profiles characterization, but focused on both the academic background and the professional background. Besides, our approach uses a really higher number of profiles instead of the 175 used in [29]. Clustering techniques were also applied in [30] with the aim of detecting groups or communities in this social network. They have also used as dataset a small group of 300 profiles. Finally, in [31] authors focus on detecting spammer's profiles. Their work on 750 profiles concludes a third of the profiles were identified as spammers.

For this study, we have collected more than 5.7 million LinkedIn profiles by scraping its public members' directory. Then, and after cleaning the obtained data set, we have classified the profiles according to their educational background into 5 categories (PhD, master, bachelor, secondary, and others) and by considering the 4 levels of the UNESCO educational classification. NLP techniques were applied for this task but also for clustering the professional background of the profiles belonging to the PhD category. In this context, we have applied the well-known K-means algorithm conjunctly with the elbow method as a heuristic to determine the appropriate k-value. Finally, and for each cluster, we have generated the tag cloud associated to the professional description of the profiles. This characterization enables us to provide more insights about the professional groups of an educational category.

Finally, and having established the former groups of educationally/professionally similar groups, we are currently working on given answers to the following questions: to what extent does educational background impact in the professional success? In how much time does this impact get its maximum level? Besides, and with the availability of temporal information in our data set (dates related to the job experience but also to the periods of studies or degrees), the application of predictive techniques is one of our highest priorities in order to provide career path recommendations according to the job market needs.

## ACKNOWLEDGMENT

This work is funded by Spanish Ministry of Economy and Competitiveness under the National Science Program (TIN2010-20797 & TEC2013-47665-C4-3-R); the European Regional Development Fund (ERDF) and the Galician Regional Government under agreement for funding the Atlantic Research Center for Information and Communication Technologies (AtlantTIC); and the Spanish Government and the European Regional Development Fund (ERDF) under project TACTICA. This work is also partially funded by the European Commission under the Erasmus Mundus GreenIT project (3772227-1-2012-ES-ERA MUNDUS-EMA21). The authors also thank GRADIANT for its computing support.

## Authors

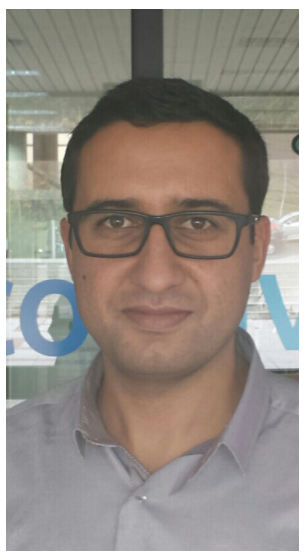

**Kais Dai** is a Ph.D. Student in Information and Communication Technologies at the University of Vigo (Spain) and member of the I&C Lab. (AtlanTIC Research Center) since 2013. His research interests are focused on Social Data Mining, Learning Analytics and Optimization Techniques. Kais obtained his master’s degree in *New Technologies of Dedicated Computing Systems* from the National Engineering School of Sfax (Tunisia) in 2012. He worked on several IT projects mainly with the UNIDO (United Nations Industrial Development Organization).

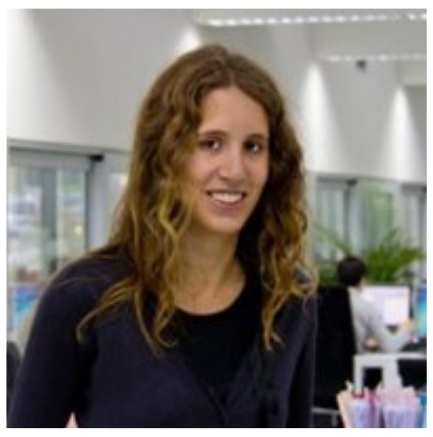

**Celia González Nespereira** is a PhD Student at the Department of Telematics Engineering of the University of Vigo. She received the Telecommunications Engineer degree from the University of Vigo in 2012 and the Master in Telematics Engineering from the same university in 2013. Celia worked as a R&D engineer at Gradiant, where she developed some projects related with media distribution protocols, web interfaces and mobile applications. In 2014 she joined the I&C Lab to work in the field of data analytics (social mining, learning analytics, etc.).

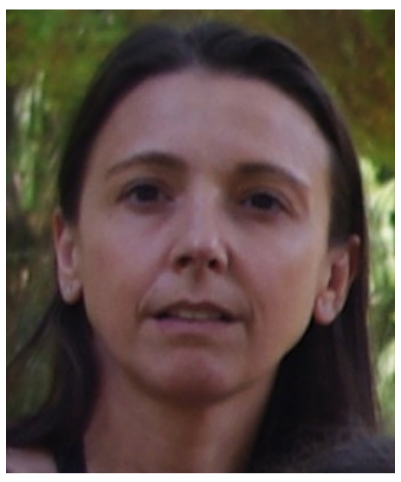

**Ana Fernández Vilas** is Associate Professor at the Department of Telematics Engineering of the University of Vigo and researcher in the Information & Computing Laboratory (AtlantTIC Research Center). She received her PhD in Computer Science from the University of Vigo in 2002. Her research activity at I&CLab focuses on Semantic-Social Intelligence & data mining as well as their application to Ubiquitous Computing and Sensor Web; Urban Planning & Learning analytics. Also she is involved in several mobility & cooperation projects with North Africa& Western Balkans.

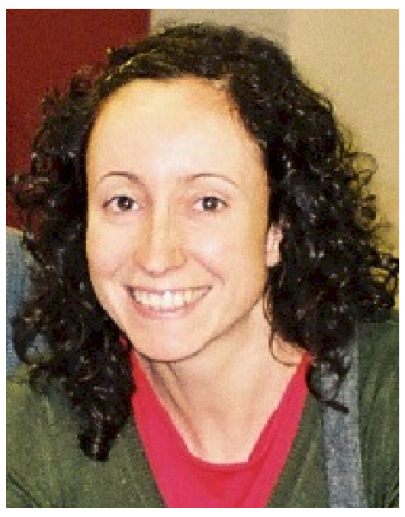

**Rebeca P. Díaz Redondo** (Sarria, 1974) is a Telecommunications Engineer from the University of Vigo (1997) with a PhD in Telecommunications Engineering from the same university (2002) and an Associate Professor at the Telematics Engineering Department at the University of Vigo. Her research interests have evolved from the application of semantic reasoning techniques in the field of Interactive Digital TV applications (like t-learning, t-goverment or TV-recommender systems) to other content characterization techniques based on collaborative labelling. She currently works on applying social mining and data analysis techniques to characterize the behavior of users and communities to design solutions in learning, smart cities and business areas.